\documentclass[a4paper,12pt]{article}

\usepackage{indentfirst}

\usepackage{amsfonts,amsmath,amssymb,bm,a4wide,graphicx,cite,makeidx,multicol}

\begin{document}

\title{Dirac and Majorana neutrinos in matter}

\author{Alexander Grigoriev\footnote{ax.grigoriev@mail.ru}, \
        Alexander Studenikin \footnote{studenik@srd.sinp.msu.ru },
   \\
   \small {\it Department of Theoretical Physics,}
   \\
   \small {\it Moscow State University,}
   \\
   \small {\it 119992 Moscow,  Russia }
   \\
   Alexei Ternov \footnote {a\_ternov@mail.ru}
   \\
   \small{\it Department of Theoretical Physics,}
   \\
   \small {\it Moscow Institute for Physics and Technology,}
   \\
   \small {\it 141700 Dolgoprudny, Russia }}

\date{}
\maketitle

\sloppy

\begin{abstract}
We consider the matter effects on neutrinos moving in background on the basis of the corresponding quantum wave
equations. Both Dirac and Majorana neutrino cases are discussed. The effects for Dirac neutrino reflection and trapping
as well as neutrino-antineutrino annihilation and $\nu\overline{\nu}$ pair creation in matter at the interface between
two media with different densities are considered. The spin light of neutrino in matter is also discussed.
\end{abstract}

\maketitle

\section{Introduction}

In the present paper we discuss a recently developed quantum approach to description of the various processes involving
neutrino in matter. In Section 2 we consider the both Dirac and Majorana neutrinos and derive the correspondent wave
equations. The obtained energy spectrum for the Dirac neutrino enables us to consider the effects of neutrino
reflection and trapping, as well as neutrino-antineutrino annihilation and $\nu\overline{\nu}$ pair creation in matter
at the interface between two media with different densities. It also reproduces the expressions for the flavour
\cite{WolPRD78MikSmiYF85} and spin-flavour \cite{Akh88,LimMar88} oscillation probabilities in matter. The quantum
theory of the spin light of neutrino in matter is presented in Section 3. In Section 4 we summarize the main results
obtained.

\section{Neutrino energy in matter}

In the series of our papers \cite{EgoLobStuPLB00-LobStuPLB04} we elaborated the quasi-classical theory of neutrino
interaction with matter. In particular, the generalized Bargmann-Michel-Telegdi equation, which was used for
description of the neutrino spin evolution with account of matter influence, was proposed. In the framework of this
approach, we predicted the existence of a new mechanism of electromagnetic radiation of a neutrino moving in matter,
which we termed "spin light of neutrino". However, it is clear, that the quasi-classical approach to the description of
this phenomenon is far not complete, because the discussed phenomenon has a quantum nature. Therefore, it is important
to elaborate the quantum approach to the description of the matter effect on neutrinos.

\subsection{Dirac neutrino}

To develop the quantum treatment of the matter effect on neutrino we propose a modified Dirac equation for the neutrino
wave function in matter. Consider the case of matter composed of electrons, neutrons and protons and assume that the
interaction of a neutrino with matter is governed by the Standard Model supplied with the right-handed singlet
neutrino. We also assume at first that the neutrino is of Dirac nature. The case of the Majorana neutrino will also be
considered in Section~2.2. The corresponding additional term in the neutrino effective Lagrangian, coming from neutrino
interactions with background via charged and neutral currents, takes the form
\begin{equation}\label{Lag_f}
\Delta L_{eff}=-f^\mu \Big(\bar \nu \gamma_\mu {1+\gamma^5 \over 2} \nu \Big), \ \ f^\mu=\sqrt2G_F
\sum\limits_{f=e,p,n} j^{\mu}_{f}q^{(1)}_{f}+\lambda^{\mu}_{f}q^{(2)}_{f},
\end{equation}
where
\begin{equation}\label{q_f}
  q^{(1)}_{f}=
(I_{3L}^{(f)}-2Q^{(f)}\sin^{2}\theta_{W}+\delta_{ef}), \ \ q^{(2)}_{f}=-(I_{3L}^{(f)}+\delta_{ef}),
\ \ \delta_{ef}=\left\{
\begin{tabular}{l l}
1 & for {\it f=e}, \\
0 & for {\it f=n, p}. \\
\end{tabular}
\right.
\end{equation}
Here $I_{3L}^{(f)}$ and $Q^{(f)}$ are the third isospin component and the electric charge of the
background fermion $f=e,n,p$. The matter currents $j_{f}^{\mu}$ and the polarization vectors
$\lambda_{f}^{\mu}$ are given by
\begin{equation}
j_{f}^\mu=(n_f,n_f{\bf v}_f), \label{j}
\end{equation}
and
\begin{equation} \label{lambda}
\lambda_f^{\mu} =\Bigg(n_f ({\bm \zeta}_f {\bf v}_f ), n_f {\bm \zeta}_f \sqrt{1-v_f^2}+ {{n_f {\bf
v}_f ({\bm \zeta}_f {\bf v}_f )} \over {1+\sqrt{1- v_f^2}}}\Bigg),
\end{equation}
$\theta _{W}$ is the Weinberg angle, $n_f$, ${\bf v}_f$ and ${\bm \zeta}_f \ (0\leqslant |{\bm \zeta}_f |^2 \leqslant
1)$ denote, respectively, the number density of the background fermions $f$, the speed of the reference frame in which
the mean momentum of the fermions $f$ is zero, and the mean value of the polarization vector of the background fermions
$f$ in the above men\-tioned reference frame. A detailed discussion on the determination of these quantities can be
found in \cite{EgoLobStuPLB00-LobStuPLB04}. The additional term (\ref{Lag_f}) leads to the modified Dirac equation for
the neutrino wave function in matter
\begin{equation}\label{Dirac} \Big\{
i\gamma_{\mu}\partial^{\mu}-\frac{1}{2} \gamma_{\mu}(1+\gamma_{5})f^{\mu}-m \Big\}\Psi(x)=0.
\end{equation}
This equation with various modifications was previously used in
\cite{BerVysBerSmiPLBGiuKimLeeLamPRD92,ChaZiaPRD88,ManPRD88,NotRafNPB88,OraSemSmoPLB_89,NiePRD89,
HaxZhaPRD91,PanPLB91-PRD92,GiuKimLeeLamPRD92,WeiKiePRD97,KachPLB98,KusPosPLB02} while studying of the neutrino
dispersion relation and oscillation probabilities in matter as well as for discussion of neutrino mass generation.

For some important cases equation (\ref{Dirac}) can be solved exactly. Let us find its solutions in the simplest case
of electron neutrino moving in unpolarized matter, ${\bm \zeta}=0$, composed of electrons. We first rewrite the
equation in the Hamiltonian form
\begin{equation}\label{H_matter}
i\frac{\partial}{\partial t}\Psi({\bf r},t)=\hat H_{matt}\Psi({\bf r},t),
\end{equation}
where
\begin{equation}\label{H_G}
  \hat H_{matt}=\hat {\bm{\alpha}} \hat {\bf p} + \hat {\beta}m +
  \hat V_{matt},
\end{equation}
and the effective potential of neutrino interaction is given by
\begin{equation}\label{V_matt}
\hat V_{matt}= \frac{1}{2\sqrt{2}}(1+\gamma_{5}){\tilde {G}}_{F}n.
\end{equation}
The Hamiltonian (\ref{H_G}) commutes with the operators of the momentum $\hat {\bf p}$ and the longitudinal
polarization ${\hat{\bf \Sigma}} {\bf p}/p$, thus we can write:
\begin{equation}\label{stat_states}
   \Psi_{s}({\bf r},t)=e^{-i(  E_{\varepsilon}t-{\bf p}{\bf r})}u_{s}({\bf p},E_{\varepsilon}),
\end{equation}
where $s=\pm 1$ is the neutrino helicity, which is determined by the equation:
\begin{equation}\label{helicity}
  \frac{{\hat{\bf \Sigma}}{\bf p}}{p}
  \Psi_{s}({\bf r},t)=s\Psi_{s}({\bf r},t),
 \ \ {\hat {\bf \Sigma}}=
\left(
   \begin{array}{cc}
   {\hat {\bm \sigma}} & 0 \\
   0 & {\hat {\bm \sigma}}
   \end{array}
\right).
\end{equation}
Under the assumption that equation (\ref{Dirac}) has a non-trivial solution, we arrive at the expression for the
neutrino energy in matter in the form:
\begin{align}
  \label{Energy}
  & E_{\varepsilon}=\varepsilon{\sqrt{{\bf p}^{2}\Big(1-s\alpha \frac{m}{p}\Big)^{2}
  +m^2} +\alpha m} ,
  \\
  \label{alpha}
  & \alpha=\frac{1}{2\sqrt{2}}{\tilde G}_{F}\frac{n}{m}.
\end{align}
The quantity $\varepsilon=\pm 1$ splits the solutions into the two branches that in the limit of the vanishing matter
density, $\alpha\rightarrow 0$, reproduce the positive- and negative-frequency solutions, respectively. An important
feature here is that the neutrino energy in matter depends on the neutrino helicity state. Therefore in the
relativistic case the left-handed and the right-handed neutrinos with equal momenta have different energies. In the
general case of matter composed of electrons, neutrons and protons, the matter density parameter $\alpha$ for different
neutrinos is
\begin{equation}\label{alpha}
  \alpha_{\nu_e,\nu_\mu,\nu_\tau}=
  \frac{1}{2\sqrt{2}}\frac{G_F}{m}\Big(n_e(4\sin^2 \theta
_W+\varrho)+n_p(1-4\sin^2 \theta _W)-n_n\Big),
\end{equation}
where $\varrho=1$ for the electron neutrino, $\nu_e$ and $\varrho=-1$ for the muon and tau neutrinos.

Using the standard procedure known from many text-books on quantum mechanics, we finally arrive at the neutrino wave
function in matter:
\begin{equation}\label{wave_function}
\Psi_{\varepsilon, {\bf p},s}({\bf r},t)=\frac{e^{-i( E_{\varepsilon}t-{\bf p}{\bf
r})}}{2L^{\frac{3}{2}}}
\begin{pmatrix}{\sqrt{1+ \frac{m}{ E_{\varepsilon}-\alpha m}}}
\ \sqrt{1+s\frac{p_{3}}{p}}
\\
{s \sqrt{1+ \frac{m}{ E_{\varepsilon}-\alpha m}}} \ \sqrt{1-s\frac{p_{3}}{p}}\ \ e^{i\delta}
\\
{  s\varepsilon\sqrt{1- \frac{m}{ E_{\varepsilon}-\alpha m}}} \ \sqrt{1+s\frac{p_{3}}{p}}
\\
{\varepsilon\sqrt{1- \frac{m}{ E_{\varepsilon}-\alpha m}}} \ \ \sqrt{1-s\frac{p_{3}}{p}}\
e^{i\delta}
\end{pmatrix},
\end{equation}
where $\delta=\arctan{p_2/p_1}$.

Let us focus upon the consideration of the energy spectrum (\ref{Energy}) in some detail. As usual, we treat the value
of the energy spectrum (\ref{Energy}), corresponding to $\varepsilon =-1$, as the antineutrino energy. Thus, for the
fixed momentum ${\bf p}$ we have four different energy values, that correspond to four possible combinations of $s$ and
$\varepsilon$ values
\begin{equation}\label{Energy_nu}
  E^{s=+1}={\sqrt{{\bf p}^{2}\Big(1-\alpha \frac{m}{p}\Big)^{2}
  +m^2} +\alpha m}, \ \ \
 E^{s=-1}={\sqrt{{\bf p}^{2}\Big(1+\alpha \frac{m}{p}\Big)^{2}
  +m^2} +\alpha m},
\end{equation}
\begin{equation}\label{Energy_anti_nu}
  {\tilde E}^{s=+1}={\sqrt{{\bf p}^{2}
  \Big(1-\alpha \frac{m}{p}\Big)^{2}
  +m^2} -\alpha m}, \ \ \
  {\tilde E}^{s=-1}={\sqrt{{\bf p}^{2}
  \Big(1+\alpha \frac{m}{p}\Big)^{2}
  +m^2} -\alpha m}.
\end{equation}
The first pair is for the neutrino quantum states of positive and negative chiralities and the
second one is for the antineutrino quantum states.

The analysis of the obtained energy spectrum (\ref{Energy_nu}), (\ref{Energy_anti_nu}) enables us to predict some
interesting phenomena that may appear at the interface of two media with different densities and, in particular, on the
motion of the neutrino at the interface between matter and vacuum. Indeed, as it follows from (\ref{Energy_nu}) and
(\ref{Energy_anti_nu}), the band-gap for neutrino and antineutrino in matter is displaced with respect to the vacuum
case in neutrino mass and is determined by the condition $\alpha m- m\leq E<\alpha m + m$. For instance, let the case
when there is no band-gap overlapping be realized (this is possible for $\alpha>2$). This situation is illustrated on
Fig.1.
\begin{figure}[h]\label{Reflaction}
\begin{center}
\includegraphics[width=0.7\textwidth]{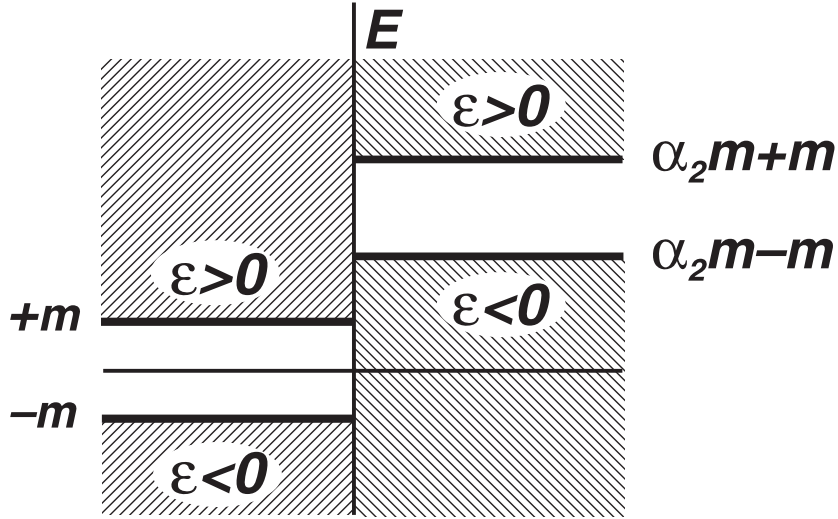}\\
\parbox{0.75\textwidth}{\caption{The interface between the vacuum (left-hand side of the picture)
    and the matter (right-hand side of the picture) with the corresponding
    neutrino band-gaps shown. The parameter $\alpha >2$. }}
\end{center}
\end{figure}
Here we have several possibilities. First, consider a neutrino moving in the vacuum towards the interface with energy
that falls into the band-gap region in matter. In this case the neutrino has no chance to survive in the matter and
thus will be reflected. The same situation will occur for the antineutrino, moving in the matter with energy falling
into the band-gap in the vacuum. In this case, there will be effect of antineutrino trapping by matter. The second
possibility is realized when the energy of neutrino in the vacuum or antineutrino in the medium falls into the region
between the two band-gaps. In this case the effects of the neutrino-antineutrino annihilation or pair creation may
occur.

\subsection{Majorana neutrino}

We have considered so far the case of the Dirac neutrino. Now we discuss the case of the Majorana neutrino. For the
Majorana neutrino, we derive the following contribution to the effective Lagrangian accounting for the interaction with
the background medium
\begin{equation}\label{Lag_f_Majorana}
\Delta L_{eff}=-f^\mu (\bar \nu \gamma_\mu \gamma^5 \nu ),
\end{equation}
which leads to the Dirac equation
\begin{equation}\label{Dirac_Majorana} \Big\{
i\gamma_{\mu}\partial^{\mu}-\gamma_{\mu}\gamma_{5}f^{\mu}-m \Big\}\Psi(x)=0.
\end{equation}
This equation differs from the one obtained in the Dirac case by doubling of the interaction term and lack of the
vector part. The corresponding energy spectrum for the equation (\ref{Dirac_Majorana}) follows straightforwardly:
\begin{equation}\label{Energy_Majorana}
  E_{\varepsilon}=\varepsilon{\sqrt{{\bf p}^{2}\Big(1-2s\alpha \frac{m}{p}\Big)^{2}
  +m^2}}.
\end{equation}
From this expression it is clear that the energy of the Majorana neutrino has its minimal value equal to the neutrino
mass, $E =m$. This means that no effects are anticipated with the Majorana neutrino such as those that the Dirac
neutrino has at the two-media interface and which are discussed above. So that, in particular, there is no Majorana
neutrino trapping and reflection by matter. It should be noted that the Majorana neutrino spectrum in matter was
discussed previously in \cite{WeiKiePRD97,PanPLB91-PRD92,GiuKimLeeLamPRD92}

\subsection{Flavour neutrino energy difference in matter}

In the conclusion to this section, we should like to note that the obtained spectra for the flavor neutrinos of
different chiralities in the presence of matter accounts properly for the effect of resonance amplification of neutrino
flavor and spin oscillations. In order to show that, we expand the expression for the relativistic electron and muon
neutrino energy, (\ref{Energy}) for Dirac case or (\ref{Energy_Majorana}) for Majorana case, over the large $p$
\begin{equation}E_{\nu_e, \nu_{\mu}}^{s=-1}\approx E_0
+ 2\alpha_{\nu_e, \nu_{\mu}} m_{\nu_e, \nu_{\mu}},
\end{equation}
where $m_{\nu_{e}}$ and $m_{\nu_{\mu}}$ are electron and muon neutrino masses. Then the energy
difference for the two active flavour neutrinos will be
\begin{equation}\Delta E =
E_{\nu_e}^{s=-1}-E_{\nu_{\mu}}^{s=-1} = \sqrt2 G_F n_e.
\end{equation}
Analogously, considering the spin-flavour oscillations $\nu_{e_L}\rightleftarrows \nu_{\mu_R}$, for
the corresponding energy difference we find:
\begin{equation}\Delta E =
E_{\nu_e}^{s=-1}-E_{\nu_{\mu}}^{s=+1} = \sqrt2 G_F \big(n_e-{1\over 2}n_n\big).
\end{equation}
These equations enable one to get the expressions for the neutrino flavour and spin-flavour oscillation probabilities
and their resonance dependances on the matter density in complete agreement with the results of
\cite{WolPRD78MikSmiYF85,Akh88,LimMar88}.

\section{Neutrino spin light in matter}

The neutrino spin light phenomenon arises in the matter-induced quantum transitions between the two neutrino states
with different helicities due to the neutrino magnetic moment interactions with photons. In this section, we give the
quantum theory of the effect that is based on an approach similar to the Furry representation in the quantum
electrodynamics, which is widely used for description of the electromagnetic interactions in the presence of external
electromagnetic fields. Within this approach, the Feynman diagram of the process under consideration is given by the
Fig.2, with the neutrino initial $\psi_{i}$ and final $\psi_{f}$ states, described by "broad lines", that account for
the neutrino interaction with matter.
\begin{figure}[h]\label{diagram}
\begin{center}
{  \includegraphics[scale=.7]{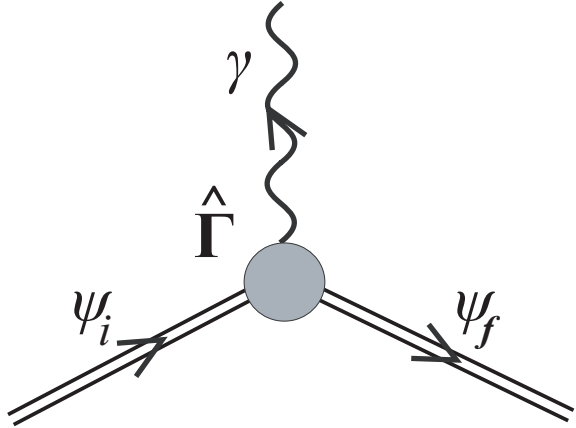}}
    \caption{
    The $SL\nu$ radiation diagram. }
  \end{center}
\end{figure}
The corresponding amplitude is given by
\begin{equation}\label{amplitude}
\begin{array}{c} \displaystyle
  S_{f i}=-\mu \sqrt{4\pi}\int d^{4} x {\bar \psi}_{f}(x)
  ({\hat {\bf \Gamma}}{\bf e}^{*})\frac{e^{ikx}}{\sqrt{2\omega L^{3}}}
   \psi_{i}(x),
   \\
   \\
   \hat {\bf \Gamma}=i\omega\big\{\big[{\bf \Sigma} \times
  {\bm \varkappa}\big]+i\gamma^{5}{\bf \Sigma}\big\},
\end{array}
\end{equation}
where $\mu$ is the neutrino magnetic moment, $k^{\mu}=(\omega,{\bf k})$ and ${\bf e}^{*}$ are the
photon momentum and polarization vectors, ${\bm \varkappa}={\bf k}/{\omega}$ is the unit vector
pointing in the direction of the emitted photon propagation. Integration over the time and spatial
coordinates gives
\begin{equation}\label{amplitude_evaluated}
   S_{f i}=
  -\mu {\sqrt {\frac {2\pi}{\omega L^{3}}}}
  ~2\pi\delta(E^{\prime}-E+\omega)\delta^3({\bf p^{\prime}}-{\bf p}+{\bf k})
  {\bar u}_{f}(E^{\prime}, {\bf p^{\prime}})({\hat {\bf \Gamma}}{\bf e}^{*})
  u_{i}(E, {\bf p}).
\end{equation}
The unprimed and primed symbols refer to initial and final states, respectively. Owing to the presence of
$\delta$-functions in the right-hand side of the equation (\ref{amplitude_evaluated}) we have the energy-momentum
conservation law for the process
\begin{equation}\label{e_m_con}
    E=E^{\prime}+\omega, \ \ \
    {\bf p}={\bf p}^{\prime}+{\bm k},
\end{equation}
from which it follows that a photon is radiated only when neutrino initial and final states are characterized by
$s_{i}=-1$ and $s_{f}=+1$, respectively. For the emitted photon energy we then obtain:
\begin{equation}\label{omega1}
\omega =\frac{2\alpha mp\left[ (E-\alpha m)-\left( p+\alpha m\right) \cos \theta \right] }{\left(
E-\alpha m-p\cos \theta \right) ^{2}-\left( \alpha m\right) ^{2}},
\end{equation}
where $\theta$ is the angle between ${\bm \varkappa}$ and the direction of the initial neutrino
propagation.

Our next step is to calculate the spin light transition rate and total radiation power, which following equations
(\ref{amplitude_evaluated}) and (\ref{omega1}) are expressed as
\begin{eqnarray}\label{Gamma}
 \Gamma &=&\mu ^2
 \int_{0}^{\pi }\frac{\omega ^{3}}{1+\tilde\beta ^{\prime
}y}S\sin \theta d\theta
\end{eqnarray}
and
\begin{equation}\label{power}
I=\mu ^2\int_{0}^{\pi }\frac{\omega ^{4}}{1+\tilde\beta ^{\prime}y}S\sin \theta d\theta,
\end{equation}
where we have denoted
\begin{equation}\label{S}
S=(\tilde\beta \tilde\beta ^{\prime }+1)(1-y\cos \theta )-(\tilde\beta +\tilde\beta ^{\prime })
(\cos \theta -y)
\end{equation}
\begin{equation}\label{beta}
\tilde \beta =\frac{p+\alpha m}{E-\alpha m}, \ \ \tilde \beta ^{\prime }=\frac{p^{\prime }-\alpha
m}{E^{\prime }-\alpha m}, \ \ \ E^{\prime }=E-\omega , \ \ \ p^{\prime }=K\omega -p,
\end{equation}
\begin{equation}
y=\frac{\omega -p\cos \theta }{p^{\prime }}, \ \ K=\frac{E-\alpha m-p\cos \theta }{\alpha m}.
\end{equation}
Performing the integration in (\ref{Gamma}) and (\ref{power}) we obtain
\begin{eqnarray}\label{W_fi}
\Gamma &=&\frac{1}{2\left( E-p\right) ^{2}\left( E+p-2\alpha m\right) ^{2}\left( E-\alpha m\right)
p^{2}} \notag
\\
&&\times \left\{ \left( E^{2}-p^{2}\right) ^{2}\left( p^{2}-6\alpha ^{2}m^{2}+6E\alpha
m-3E^{2}\right) \left( \left( E-2\alpha m\right) ^{2}-p^{2}\right) ^{2}\right. \notag
\\
&&\times \ln \left[ \frac{\left( E+p\right) \left( E-p-2\alpha m\right) }{\left( E-p\right) \left(
E+p-2\alpha m\right) }\right] +4\alpha mp\left[ 16\alpha ^{5}m^{5}E\left( 3E^{2}-5p^{2}\right)
\right.\notag
\\
&&-8\alpha ^{4}m^{4}\left( 15E^{4}-24E^{2}p^{2}+p^{4}\right) +4\alpha ^{3}m^{3}E\left(
33E^{4}-58E^{2}p^{2}+17p^{4}\right) \notag
\\
&&-2\alpha ^{2}m^{2}\left( 39E^{2}-p^{2}\right) \left( E^{2}-p^{2}\right) ^{2}+12\alpha mE\left(
2E^{2}-p^{2}\right) \left( E^{2}-p^{2}\right) ^{2} \notag
\\
&&-\left. \left. \left( 3E^{2}-p^{2}\right) \left( E^{2}-p^{2}\right) ^{3} \right] \right\}
\end{eqnarray}
and
\begin{align}
  I = & \frac{5}{2\left( E-p\right)^{3}\left( E+p-2\alpha m\right)^{3}p^{2}}
\times \
   \Bigg\{
      (E+p)^2(E-m)^3(E+p-2\alpha m)^3  \notag
\\
      & \times (E-p-2\alpha m)^2\Big(2\alpha^2 m^2-2\alpha m(E+\frac{1}{5}p)+E^2-\frac{3}{5}p^2\Big) \notag
\\
      & \times \ln
         \left(
             \frac{(2\alpha m-p-E)(E-p)}{(2\alpha m+p-E)(E+p)}
         \right)\notag
\\
   - & 4\alpha mp
          \left(32\alpha^6m^6
             \Big(
                E^4-pE^3-\frac{5}{3}p^2E^2 \right.+
                \frac{5}{3}p^3E+\frac{8}{15}p^4
             \Big)\notag
\\
 - & 96\alpha^5m^5
             \Big(
               E^5-\frac{23}{30}pE^4-\frac{83}{45}p^2E^3+
                \frac{11}{9}p^3E^2+\frac{38}{45}p^4E-\frac{1}{10}p^5
             \Big) \notag
\\
         + & 128\alpha^4m^4
             \Big(
                E^6-\frac{47}{80}pE^5-\frac{511}{240}p^2E^4 
                +\frac{127}{120}p^3E^3+
                \frac{157}{120}p^4E^2
                -\frac{89}{240}p^5E-\frac{7}{48}p^6
             \Big)
\\        - & 96(E^2-p^2)\alpha^3m^3
             \Big(
                E^5-\frac{53}{120}pE^4
                -\frac{3}{2}p^2E^3+\frac{89}{180}p^3E^2
                +\frac{47}{90}p^4E-\frac{19}{360}p^5
            \Big) \notag
\\        + & 42(E^2-p^2)^2\alpha^2m^2
             \Big(
                E^4-\frac{32}{105}pE^3
                -\frac{314}{315}p^2E^2+
                \frac{4}{21}p^3E+\frac{17}{105}p^3
             \Big) \notag
\\        - & 10\alpha m(E^2-p^2)^3
             \Big(
                E^3-\frac{4}{25}pE^2
                -\frac{17}{25}p^2E+\frac{2}{25}p^3
             \Big) \notag
\\        + & \left.(E^2-p^2)^4
             \Big(
                E^2-\frac{3}{5}p^2
             \Big)
          \right)
   \Bigg\},\notag
\end{align}
respectively. As it follows from these expressions, the $SL\nu$ rate and total power are rather complicated functions
of the neutrino momentum $p$, mass $m$, and the matter density parameter $\alpha$. It makes sense to examine them for
different limiting cases. In the relativistic case $p\gg m$, we get
\begin{equation}\label{p_gg}
\Gamma = \left\{
  \begin{tabular}{c}
  \ $\frac{64}{3} \mu ^2 \alpha ^3 p^2 m,$ \\
  \ $4 \mu ^2 \alpha ^2 m^2 p$, \\
  \ $4 \mu ^2 \alpha ^3 m^3$,
  \end{tabular}
\right. \ \ I= \left\{
  \begin{tabular}{cc}
  \ $\frac{128}{3}\mu ^{2}\alpha ^{4}p^{4},$ &
  \ \ \ for {$\alpha \ll \frac{m}{p},$ } \\
  \ $\frac{4}{3} \mu ^2 \alpha ^2 m^2 p^2$, & \ \ \ \ \ \ \ \  { for
  $ \frac{m}{p} \ll \alpha \ll \frac{p}{m},$} \\
\ $4 \mu ^2 \alpha ^4 m^4$, & \ { for
  $ \alpha \gg \frac{p}{m}, $}
  \end{tabular}
\right.
\end{equation}
and in the opposite case, $p\ll m$, we have
\begin{equation}\label{p_ll}
\Gamma = \left\{
  \begin{tabular}{c}
  \ $\frac{64}{3} \mu ^2 \alpha ^3 p^3,$ \\
  \ $\frac{512}{5} \mu ^2 \alpha ^6 p^3$,\\
  \ $4 \mu ^2 \alpha ^3 m^3$,
  \end{tabular}
  \ \ \
I = \left\{
  \begin{tabular}{cc}
  \ $\frac{128}{3} \mu ^2 \alpha ^4 p^4,$ &
  \ \ \ \ \ for {$\alpha \ll 1,$ } \\
  \ $\frac{1024}{3} \mu ^2 \alpha ^8 p^4$, & \ \ \ \ \ \ \ \ \ \ { for
  $ 1 \ll \alpha \ll \frac{m}{p},$} \\
  \ $4 \mu ^2 \alpha ^4 m^4$, & \ \ \ \ { for
  $ \alpha \gg \frac{m}{p}. $}
  \end{tabular}
\right. \right.
\end{equation}
It is interesting to observe that, in the limit of a very high matter density, the rate and the radiation power are
determined by the background matter density only. Note that the obtained $SL\nu$ rate and radiation power for $p\gg m$
and $\alpha \gg \frac{m}{p}$ are in agreement with \cite{Lob_hep_ph_0411342}.

One of the important aspects of the $SL\nu$ radiation is its polarization properties, which may be
important in experimental identification of the radiation. We will be interested here in the
circular polarization of the $SL\nu$ photons. We introduce the two orthogonal vectors
\begin{equation}\label{circ_pol}
  {\bf e}_{l}=\frac{1}{\sqrt 2}({\bf e}_{1}+il{\bf e}_{2}),
\end{equation}
that describe the two $SL\nu$ photon circular polarizations ($l=\pm 1$ correspond to the right and
left photon circular polarizations, respectively), via the two linear photon polarization vectors,
defined as
\begin{equation}\label{e_12}
  {\bf e}_1= \frac{[{\bm \varkappa}\times {\bf j}]}
  {\sqrt{1-({\bm \varkappa}{\bf j})^{2}}}, \ \
  {\bf e}_2= \frac{{\bm \varkappa}({\bm \varkappa}{\bf j})-{\bf j}}
  {\sqrt{1-({\bm \varkappa}{\bf j})^{2}}},
\end{equation}
where $\bf j$ is the unit vector pointing in the direction of the initial neutrino propagation.
Then for the radiation power of the circular-polarized photons we get
\begin{equation}
I^{\left( l\right) }=\mu ^2\int_{0}^{\pi }\frac{\omega ^{4}}{1+\beta ^{\prime }y}S_{l}\sin \theta
d\theta ,
\end{equation}
where
\begin{equation}
S_{l}=\frac{1}{2}\left( 1+l\beta ^{\prime }\right) \left( 1+l\beta \right) \left( 1-l\cos \theta
\right) \left( 1+ly\right).
\end{equation}
The most important case is realized for rather dense matter and a relativistic neutrino, $\alpha \gg \frac{m}{p}$ for
$p\gg m $ and $\alpha \gg 1$ for $p\ll m$, when the main contribution to the radiation power comes from the
right-polarized photons: $I^{\left( +1\right) } \simeq I$, $I^{\left( -1\right) } \simeq 0$.
So we conclude that, in a dense matter, the $SL\nu$ photons are emitted with nearly total right-circular polarization.
If the density parameter changes sign, then the emitted photons will exhibit the left-circular polarization.

At the end of this section, we should like to make a note on the possibility of the $SL\nu$ radiation by a Majorana
neutrino. Obviously, owing to the absence of a magnetic moment, such radiation is not expected in this case. However,
having the two neutrinos of different flavour, it is possible to have an analogous effect via the transition magnetic
moment, which Majorana neutrinos may possess.

\section{Summary}

In conclusion, we would like to note the following:

1) Propagation of the Dirac and Majorana neutrinos through the moving and polarized matter can be described by the
modified Dirac equations, given by (\ref{Dirac}) for Dirac and by (\ref{Dirac_Majorana}) for Majorana particles.

2) In the Dirac case, the effects of neutrino reflection and trapping, as well as neutrino-antineutrino annihilation
and pair creation can occur at the interface between the two media of different densities.

3) While moving in matter, the Dirac neutrino emits $SL\nu$ radiation due to its nonzero magnetic moment; this effect
originates from the induced by matter quantum transitions between the two neutrino states with different helicities.

4) Majorana neutrinos do not radiate $SL\nu$ because the magnetic moment of this particle is zero; however, the $SL\nu$
in matter can be produced in the case of two Majorana neutrinos of different flavour if the transition magnetic moment
is nonzero.

5) The $SL\nu$ radiation exhibits a crucial dependance on the matter density; for high densities, the $SL\nu$ rate and
radiation power are determined by the background matter density only, in this case the $SL\nu$ has almost total
right-circular polarization (for positive value of the matter density parameter $\alpha$).

6) The matter effects on neutrinos under consideration may have important consequences in astrophysics and cosmology.

7) The presented in this paper method can be expanded for description of the electron in matter; on this stage we
predict the analogous to $SL{\nu}$ phenomenon of "spin light of electron" and develop the quantum theory of this effect
\cite{StuJPA06-GriShinkStuTernTro12LomCon06}.

One of the authors (A.S.) is thankful to the organizers of the 5th International Conference on
Non-Accelerator New Physics for hospitality.


\begin{thebibliography}{99}

\bibitem{WolPRD78MikSmiYF85} L.~Wolfenstein, Phys. Rev. D {\bf 17}, 2369 (1978); S.~Mikheyev, A.~Smirnov, Sov. J. Nucl. Phys. {\bf 42} 913 (1985).

\bibitem{Akh88} E.~Akhmedov, Phys. Lett. B {\bf 213} 64 (1988).

\bibitem{LimMar88} C.-S.~Lim, W.~Marciano, Phys. Rev. D {\bf 37} 1368 (1988).




\bibitem{EgoLobStuPLB00-LobStuPLB04}
 A.~Egorov, A.~Lobanov and A.~Studenikin, Phys. Lett. B {\bf 491} 137
(2000); A.~Lobanov, A.~Studenikin, Phys. Lett. B {\bf 515} 94 (2001);
M.~Dvornikov, A.~Studenikin, JHEP {\bf 09} 016 (2002);  A.~Lobanov, A.~Studenikin, Phys. Lett. B {\bf 564} 27 (2003);
A.~Lobanov, A.~Studenikin, Phys. Lett. B {\bf 601} 171 (2004); M.~Dvornikov, A.~Grigoriev and A.~Studenikin, Int. J.
Mod. Phys. D {\bf 14} 309 (2005); A.~Studenikin, Nucl. Phys. B (Proc.~Suppl) {\bf 143} 570 (2005).

\bibitem{StuTerPLB05} A.~Studenikin, A.~Ternov, Phys. Lett. B {\bf 608} 107 (2005)~//
hep-ph/0410297~// hep-ph/0412408.

\bibitem{StuTerQUARKS_04_0410296} A.~Studenikin, A.~Ternov, in {\it
Proc. of the 13th Int. Seminar on High Energy Physics "Quarks-2004"}, Ed. by ~D.~Levkov, ~V.~Matveev and ~V.~Rubakov
(Publ. Department of Institute of Nuclear Physics RAS, Moscow, 2004)~// hep-ph/0410296; A.~Grigoriev, A.~Studenikin and
A.~Ternov, in {\it "Particle Physics in Laboratory, Space and Universe", Procceedings of the 11th Lomonosov Conference
on Elementary Particle Physics, Moscow, 2003,} Ed. by A.~Studenikin (World Sc., Singapore, 2005), p. 55~//
hep-ph/0502210.

\bibitem{GriStuTerCOSMION04_hep_ph0502231} A.~Grigoriev,
A.~Studenikin and A.~Ternov, Grav. \& Cosm. {\bf 11} 132 (2005).

\bibitem{GriStuTerPLB_05} A.~Grigoriev, A.~Studenikin and A.~Ternov, Phys. Lett. B {\bf 622} 199 (2005).

\bibitem{BerVysBerSmiPLBGiuKimLeeLamPRD92}Z.Berezhiani, M.~Vysotsky,
Phys. Lett. B {\bf 199} 281 (1987); Z.~Berezhiani, A.~Smirnov, Phys. Lett. B {\bf 220} 279 (1989).

\bibitem{ChaZiaPRD88} L.~N.~Chang, R.~K.~Zia, Phys. Rev. D {\bf 38} 1669 (1988).

\bibitem{ManPRD88} P.~Mannheim, Phys. Rev. D {\bf 37} 1935 (1988).

\bibitem{NotRafNPB88} D.~N\"otzold, G.~Raffelt, Nucl. Phys. B {\bf 307} 924 (1988).

\bibitem{OraSemSmoPLB_89} V.~Oraevsky, V.~Semikoz and Ya.~Smorodinsky, Phys. Lett. B {\bf 227} 255 (1989).

\bibitem{NiePRD89} J.~Nieves, Phys. Rev. D {\bf 40} 866 (1989);

\bibitem{HaxZhaPRD91} W.~Haxton, W-M.~Zhang, Phys. Rev. D {\bf 43} 2484 (1991).

\bibitem{PanPLB91-PRD92} J.~Pantaleone,
Phys. Lett. B {\bf 268} 227 (1991);
Phys. Rev. D {\bf 46} 510 (1992).

\bibitem{GiuKimLeeLamPRD92} C.~Giunti, C.~W.~Kim, U.~W.~Lee, W.~P.~Lam, Phys. Rev. D {\bf 45} 1557 (1992).

\bibitem{WeiKiePRD97} K.~Kiers, N.~Weiss,
Phys. Rev. D {\bf 56} 5776 (1997); K.~Kiers, M.~Tytgat, Phys. Rev. D {\bf 57} 5970 (1998).

\bibitem{KachPLB98} M.~Kachelriess, Phys. Lett. B {\bf 426} 89 (1998).

\bibitem{KusPosPLB02} A.~Kusenko, M.~Postma, Phys. Lett. B {\bf 545} 238 (2002).

\bibitem{Lob_hep_ph_0411342} A.~Lobanov, Dokl. Akad. Nauk Ser. Fiz. {\bf 402} 475 (2005);
hep-ph/0411342; Phys. Lett. B {\bf 619} 136 (2005).

\bibitem{StuJPA06-GriShinkStuTernTro12LomCon06} A.~Studenikin, J. Phys. A: Math.Gen, {\bf 39} 6769
(2006); A.~Grigoriev, S.~Shivkevich, A.~Studenikin, A.~Ternov, I.~Trofimov, in {\it Particle
Physics at the Year of 250th Universary of Moscow University, Proceedings of the 12th Lomonosov
Conference on Elementary Particle Physics, Moscow, 2005}, Ed. by A.I.~Studenikin (World Sci.,
Singapore, 2006), p.~73.

\end{thebibliography}
\end{document}